\documentclass[10pt, journal]{IEEEtran}
\usepackage{cite}
\usepackage{multirow}
\usepackage{multicol}
\usepackage{graphicx}
\usepackage{epsfig}
\usepackage{graphics}
\usepackage{subfigure}
\usepackage{psfrag}
\usepackage{stfloats}
\usepackage{amsmath, amssymb}
\usepackage{algorithm}
\usepackage{algorithmic}
\usepackage{array}

\ifCLASSINFOpdf
\else
\fi
\begin{document}

%
\title{Label-Specific Training Set Construction from Web Resource for Image Annotation}
%
%
%

\author{Jinhui~Tang, 
        Shuicheng Yan, 
        Tat-Seng Chua, and Ramesh Jain
\thanks{J. Tang is with the School of Computer Science and Technology, Nanjing University of Science and Technology, 210094,
Nanjing, China. (email: tangjh1981@acm.org); }
\thanks{S. Yan is with the
Department of Electrical and Computer Engineering, National University of Singapore, 117576, Singapore (email: eleyans@nus.edu.sg).}
\thanks{T.-S. Chua is with the School of Computing, National University of Singapore, 117417,
Singapore (email: chuats@comp.nus.edu.sg); }
\thanks{R. Jain is with the
Department of Information and Computer Science, University of California, Irvine, 92697, CA, USA (email: jain@ics.uci.edu).}
\thanks{Manuscript received ***, 2011.}}

%
%

\markboth{}%
{Shell \MakeLowercase{\textit{et al.}}: Bare Demo of IEEEtran.cls
for Journals}
%



\maketitle

\begin{abstract}
Recently many research efforts have been devoted to image annotation by leveraging on the associated tags/keywords of web images as training labels. A key issue to resolve is the relatively low accuracy of the tags. In this paper, we propose a novel semi-automatic framework to construct a more accurate and effective training set from these web media resources for each label that we want to learn. Experiments conducted on a real-world dataset demonstrate that the constructed training set can result in higher accuracy for image annotation.
\end{abstract}

\vspace{-3mm}
%
\IEEEpeerreviewmaketitle

\section{Introduction}
\label{sec:intro}

\begin{figure}
\centering
\includegraphics[width=0.95 \linewidth]{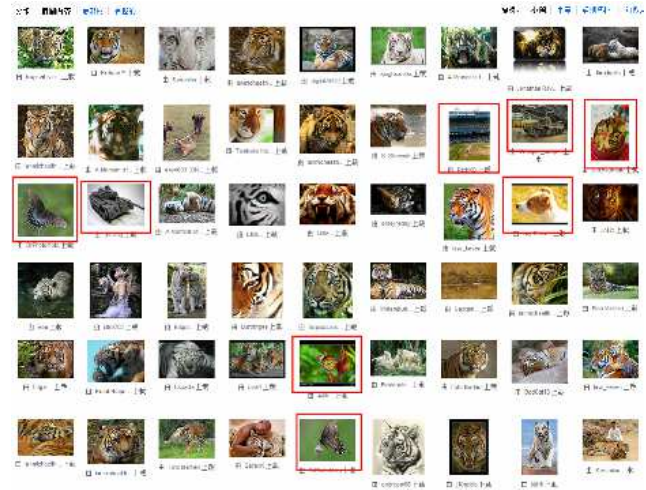}
\vspace{-0mm}
\caption{The top-retrieved results by searching ``tiger"  in Flickr. The images marked with red rectangles are incorrect results.} \label{fig:fig1}
\vspace{-0mm}
\end{figure}

Recent years have witnessed the proliferation of social media and the success of many photo-sharing websites, such as Flickr and Picasa. These websites allow users to rate and tag the shared images. For multimedia research, several methods were proposed to annotate images by leveraging on these web images and their associated tags \cite{WangXJ:PAMI08}\cite{Tang:MM09}. In these approaches, a key problem affecting the annotation performance is that the tags are too ``noisy" \cite{Tang:MM09}. Fig. \ref{fig:fig1} shows the top-retrieved results by searching ``tiger"  in Flickr. We can see that the images marked with red rectangles do not really describe the tiger we want to search. They were retrieved here due to that they are subjectively tagged with ``tiger". However, for objective image auto-annotation, we need the training labels that accurately describe the objective aspects of the visual content in the images.

Research attention has also been paid to refine the tags for the web images \cite{Xu.H:MM09}. Actually, for learning-based image annotation, we do not need to correct all the tags associated with the images, since there is an extremely huge amount of tagged images in the web. Instead, we just need to construct an effective training set for each label that we want to learn. It requires two properties for the training set of each label: (i) the constructed set should have a large enough number of images that are objectively relevant to the given label; and (ii) the samples in the constructed set should have diverse low-level features. In our scenario, since the images are crawled from a huge web resource, the second property can easily be satisfied. Thus we only need to consider how to satisfy the first property.

\begin{figure*}
\centering
\includegraphics[width=0.75 \linewidth, height=0.52 \linewidth]{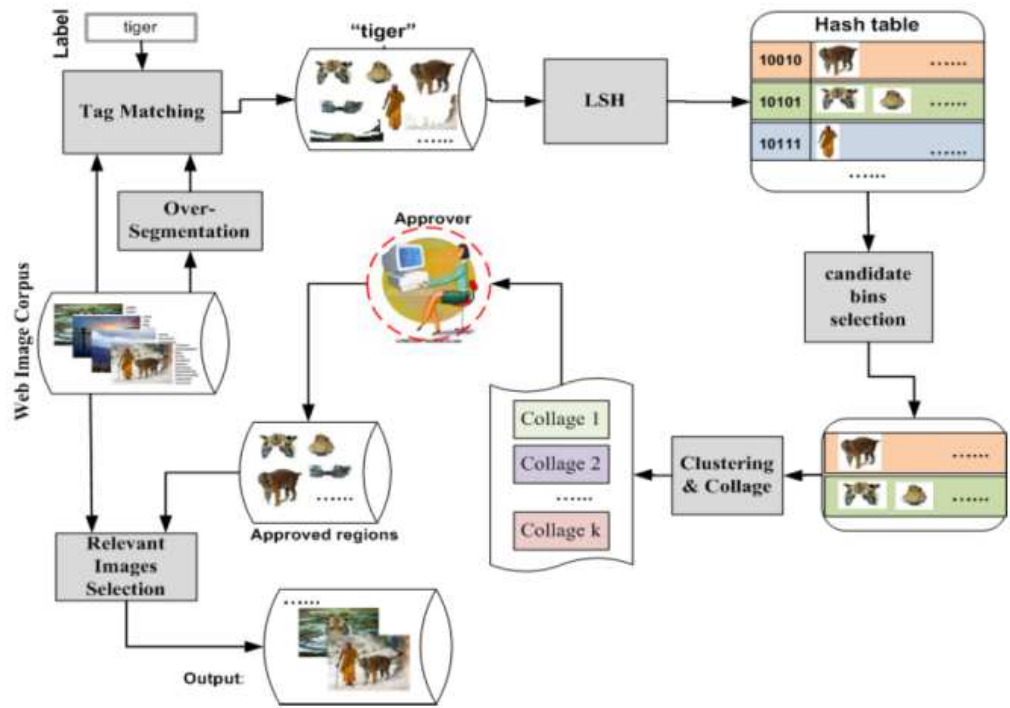}
\caption{The flowchart of the proposed framework.} \label{fig:fig3}
\end{figure*}

In this paper, we propose a novel semi-automatic framework with minimal human effort to construct an effective training set from the image-sharing sites for image annotation. Since an image is relevant to a certain label iff the label describes the content of one or more regions in this image, we first segment each image into regions. We then employ locality sensitive Hashing (LSH) \cite{LSH} to find the most possibly relevant regions (region candidates) of a given label efficiently. We further conduct simple human interactions to approve whether the clusters of region candidates are relevant to the given label. Here Hashing ensures the efficiency and the minimal human efforts guarantee the effectiveness of the proposed framework. Although there are several approaches utilizing the web resource as training data \cite{Ulges:CIVR08}\cite{Chang:TRECVID08}\cite{Setz:ICME09}\cite{LiXR:ICASSP09}\cite{LiXR:MM09s}\cite{Tang:MM09}\cite{Ulges:CVIU10}, they did not \emph{construct} more accurate training set from web resource for image annotation and search.

\section{The Proposed Framework}
\label{sec:Framework}

Fig. \ref{fig:fig3} illustrates the proposed framework for training set construction. We summarize the flowchart as follows:
\begin{enumerate}[]
\item We crawl a huge amount of images and their associated tags from image-sharing sites, such as Flickr and Picasa.
\item Each image is over-segmented into several regions and each region is described with several low-level features. Here we decided to segment the images due to two reasons: i) most of the labels attached to the images correspond to the regions but not the whole images; and ii) region features corresponding to a particular label are more consistent than the global features, thus are more suitable for fast and rough clustering with Hashing.
    \vspace{-0mm}
\item Given a label, we construct its training set as follows:\\
 (a) We map all regions of the candidate images into different bins by LSH. Here the candidate images mean the images that have the given label in their tag lists. LSH is used for efficient rough clustering since the size of web image collection is very large.\\
 (b) We take the number of regions and the variations of features in each bin as measures to select the most appropriate bins related to the given label. In our experiments, we simply select the largest bins since it is generally believed that the images, which are relevant to a certain label, will have many similar regions. Here we select the bins starting from the largest and stop when the total number of regions in all the selected bins exceeds twice the number of the candidate images. We collage the regions in each bins to request the users to approve whether the bin corresponds to a background. Here the user feedbacks are requested to remove large bins related to background regions.\\
 (c) We cluster the regions in the selected bins into different clusters. We then collage the regions in each cluster to request the users to approve whether the cluster is relevant to the given label. Here the user approvals ensure the accuracy. While they are only conducted on the clusters of most possibly relevant regions, thus many human efforts are saved. Affinity propagation \cite{AP:science07} is used in the experiments for sophisticated clustering.\\
 (d) The images including the regions in the approved clusters are selected as the relevant images.
 \vspace{-0mm}
\end{enumerate}

\begin{figure}
\centering
\includegraphics[width=0.9 \linewidth]{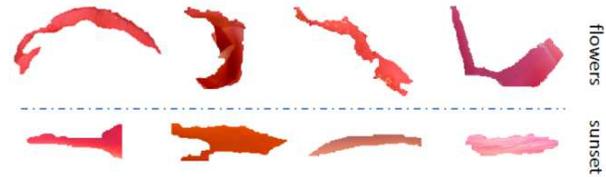}
\vspace{0mm}
\caption{Exemplary regions related to ``flowers" and ``sunset". \cite{Tang:MM10short}} \label{fig:fig4}
\end{figure}

Different from the typical applications of LSH for indexing or fast nearest neighbor search of samples, in our framework, we adapt LSH to coarse clustering. LSH is based on the simple idea that, if two points are close together, then after a ``projection" operation these two points will remain close together. It means that the Hashing may not be able assign all similar samples into the same bucket, but the samples assigned into the same bucket are generally similar. Thus it can be used for coarse clustering while it is very efficient. Here the size of the region corpus is very large and clustering them with normal methods (such as Kmeans \cite{Kmeans:PAMI02} and spectral clustering \cite{Ng01:spectral}) are quite time-consuming.

\begin{figure*}
\centering
\includegraphics[width=0.98 \linewidth]{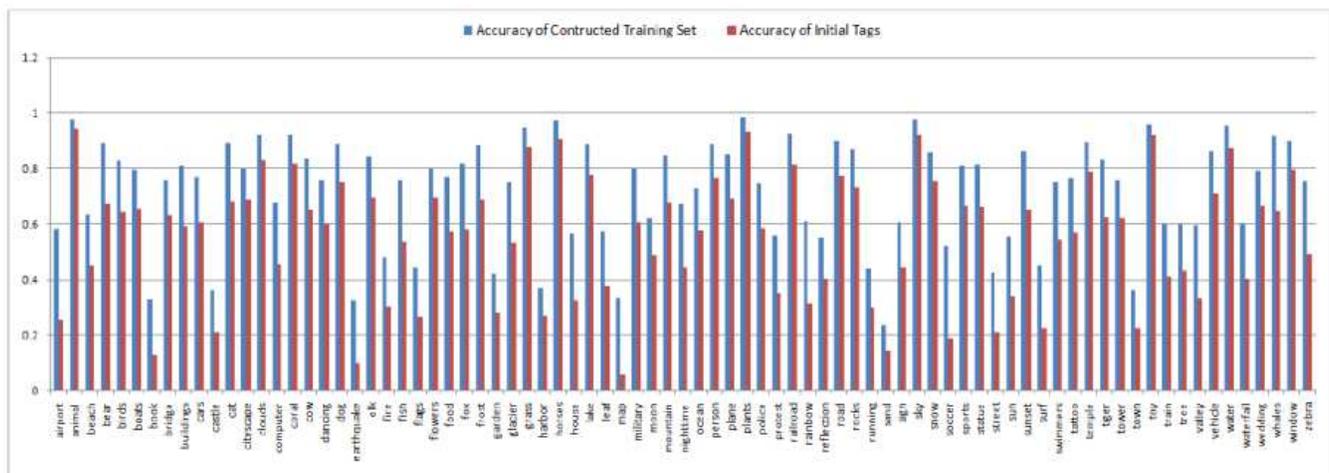}
\caption{The comparisons of the label accuracies before and after the construction.} \label{fig:accuracy}
\end{figure*}

When we present the clusters to users for approval, we need to collage the regions of a cluster into a big picture. If we only collage the regions, it is still difficult for users to identity the corresponding label. Fig. \ref{fig:fig4} illustrate an example, in which the 8 regions are respectively extracted from different images related to ``flowers" and ``sunset". We can find that the users cannot even differentiate which region corresponds to which label. According to the discussion in \cite{Jain:MM10} that ``content without context is meaningless", the contextual information should also be provided for user labeling. To this end, we collage the whole images instead of only the regions for user labeling.  Similar approach was also used in fast manual image annotation \cite{Tang:MM10short}. However, here we only need to collage all the images but do not need to differentiate the target regions and the contextual regions. Because even if the target regions are not related to the given concept, we still approve the corresponding images as relevant if the contextual regions are related to the concept.

The region features we used in this work include color correlogram, color moment, region shape and region position. Recall the exemplary regions in Fig. \ref{fig:fig4}, the human cannot even discriminate them. Thus it is impossible to discriminate them by automatic clustering only according to the region features. To handle this problem, we also utilize the global features as context for clustering. Thus the detailed process of clustering is as follows: i) for each given cluster, we first cluster it into several smaller clusters by affinity propagation on region features; and ii) we then further cluster each obtained cluster into three smaller sub-clusters by Kmeans, where we represent each region by its corresponding image features. We use Kmeans but not affinity propagation here because we need to control the cluster number. We set the cluster number to three because it is observed that in most cases the similar regions of a given region correspond to at most three labels. The global image features we used here are color histogram and edge direction histogram.

The framework can only be used to construct the positive part of the training set for each label.  We randomly sample the irrelevant images from the rest of the original image pool to form the negative part.

\section{Experimental Evaluation}
To evaluate the proposed framework for training set construction, we conduct experiments on a real-world dataset NUS-WIDE \cite{nus-wide-civr09} on 81 labels. The dataset is divided into two parts: development part, which contains 161,789 images, and testing part, which contains 107,859 images. We construct the training set from the development part.

We also use the constructed training set to annotate the images in the testing part by using the simple \emph{k}-nearest neighbors (\emph{k}-NN) method on the 81 labels. For comparison, we set up a baseline method also using \emph{k}-NN but without using the constructed training set; instead, the positive samples are sampled from the candidate images in the development part with the same number as the constructed set. For both methods, the negative samples are randomly sampled from the rest of images in the development part. And for both approaches, we run the experiments three times and average the results.

\begin{figure*}
\centering
\includegraphics[width=0.98 \linewidth]{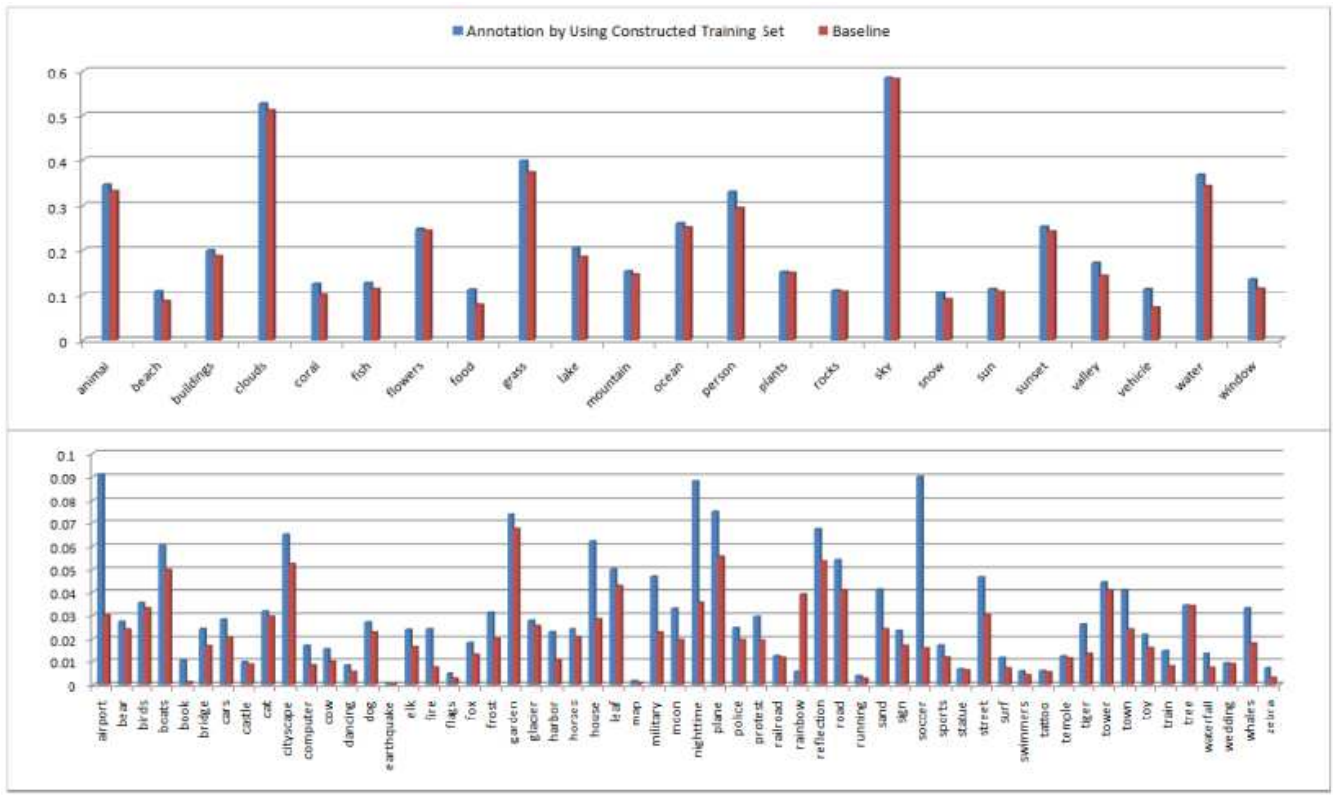}
\caption{The comparisons of the average precisions of learning with the constructed set and the baselines.} \label{fig:fig5}
\end{figure*}

We compare the accuracies of training labels before and after the construction in Fig. \ref{fig:accuracy}. We can see that after the training set construction, the accuracies of the training labels improve significantly compared to the initial tags, while the accuracies of several labels even approaches 100\%. Table \ref{tb:tb1} gives the quantitative analysis of the constructed training set. The average number of images in the constructed set for each label is 457, while the average construction rate (the number of images in the constructed training set divides the number of candidate images) is 31.8\%.  After the construction, the mean\footnote{\tiny Here ``mean" is equal to ``average". We use it to differentiate with ``average precision", which has a specific definition \cite{TRECVID_measure}.} precision of the training labels improves from 54.2\% to 75.1\%. Fig. \ref{fig:fig5} compares the average precisions (AP) \cite{TRECVID_measure} of the annotation results obtained by \emph{k}-NN using the constructed set versus the baseline results. The mean average precision (MAP) on the 81 labels is 0.087, which has an improvement of 19.2\% over that of the baseline. From these results, we can see that the proposed framework is able to construct more accurate and effective training sets for image annotation, especially on those labels that are difficult to learn (bottom part of Fig. \ref{fig:fig5}). However, we can also see the average number of approvals by users is 37 for each label. Thus there is still room to improve the efficiency.

\begin{table}[!tb]
\centering \caption{Quantitative Analysis of the Constructed Training Set} \label{tb:tb1} {
\begin{tabular}{lc} \hline
\hline   Average Number of Images              &       457                   \\
         Average Number of Approvals by Users       &       37                    \\
         Average Construction Rate     &       31.8\%                \\
         Mean Precision Improvement &       54.2\%$\rightarrow$75.1\%         \\
         Test MAP Improvement       &       0.073$\rightarrow$0.087 \\
\hline
\hline
\end{tabular}}
\end{table}

\section{Conclusion and Future Work}
In this paper, we proposed a novel framework to construct a more accurate training set from the image-sharing sites for image annotation. The key-point of this work is to construct several training samples by one user feedback. Experiments conducted on the NUS-WIDE dataset had demonstrated its effectiveness. However, in this framework, more sophisticated bin selection strategies can be designed to improve the efficiency, and more representative contextual information can be incorporated into framework to improve the effectiveness. In the future work, we will work on these two points to improve the performance while reduce the human effort.

%
\bibliographystyle{abbrv}
%
%

\vfill


\end{document}